# Feasibility and performance assessment of a practical autonomous deep space navigation system based on X-ray pulsar timing

Setnam Shemar [(1,*)], George Fraser [(2, 1)], Lucy Heil [(3)], David Hindley [(1)], Adrian Martindale [(2)], Philippa Molyneux [(2)], John Pye [(2)] & Robert Warwick [(2)]

[(1)] National Physical Laboratory, Hampton Road, Teddington, Middlesex, TW11 0LW, UK

[(2)] University of Leicester, Department of Physics and Astronomy, University Road, Leicester, LE1 7RH, UK

[(3)] Anton Pannekoek Institute for Astronomy, University of Amsterdam, Postbus 94249, 1090 GE Amsterdam, The Netherlands

## *Abstract*

Shemar et al. (2016) presented results based on the output of a feasibility study for the European Space Agency (ESA) on the use of X-ray pulsars for deep space navigation, a concept often referred to as 'XNAV'. Here we describe some of the key results as well as providing additional information which includes navigation uncertainties and the potential X-ray technology that could be used. For a conventional deep space mission, an X-ray navigation system must be practical to implement as a spacecraft subsystem and to this end it must meet restrictive mass, volume and power consumption requirements. The implementation of an X-ray observatory sized instrument is unrealistic in this case. The Mercury Imaging X-ray Spectrometer (MIXS) instrument, due to be launched on the ESA/JAXA BepiColombo mission to Mercury in 2018, is an example of an instrument that may be further developed as a practical telescope for XNAV. Simulations involving different pulsar combinations and navigation strategies are used to estimate the navigation uncertainties that may be achievable using such an instrument. Possible options for future developments in terms of simpler, lower-cost Kirkpatrick-Baez optics are discussed, in addition to the principal design and development challenges that must be addressed in order to realise an operational XNAV system.

---

[1] GF deceased 18 March 2014

[*] email: setnam.shemar@npl.co.uk

# 1    Introduction

Using radio interferometric techniques, NASA's Deep Space Network (DSN) currently has the capability to position a spacecraft in deep space to an angular resolution of 1 nrad (Curkendall and Border 2013). This is equivalent to 150 m per Astronomical Unit (AU) distance of the craft from Earth in the plane of the sky (a plane that is tangential to the celestial sphere and perpendicular to the line-of-sight from a ground-station to the craft). At a distance of 30 AU, the uncertainty is 4.5 km. Along the line-of-sight to the craft, the positioning uncertainty using one-way ranging is a few metres. The angular resolution of ESA's European Space Tracking (ES-TRACK) system (see M. ButKovic paper in this Volume) is currently on the order of 10 nrad, with a commensurately higher positioning uncertainty in the plane of the sky.

The Mercury Imaging X-ray Spectrometer (MIXS) instrument due to be launched on the ESA/JAXA BepiColombo mission to Mercury in 2018 will employ low-mass optics to implement an X-ray telescope. With further development, such an instrument offers the possibility of implementing an XNAV instrument to meet the stringent mass, size and power requirements of a conventional deep space mission. Shemar et al. (2016) presented the potential navigation performance that would be enabled by such an instrument based on the expected pulse Time-Of-Arrival (TOA) measurement accuracy of one or more X-ray pulsars. Here we summarise the main results as well as providing additional information. First, we briefly describe some of the navigation concepts behind this technique and some of the potential X-ray pulsars that could be utilized. This is followed by the results from simulations of navigation and timing uncertainties. We also outline the currently available X-ray instrumentation and potential future developments that could be used in an XNAV system. We then discuss the findings and finally give the conclusions.

# 2    Navigation concepts in XNAV

### Using delta-correction measurements

This strategy is described in Sheikh et al. 2006 and Graven et al. 2008. An initial estimate of the craft position is required to within $cP/2$, where $c$ is the speed of light and $P$ is the pulse period of the pulsar, as well as an estimate of the velocity. These could be obtained using the Global Positioning System (GPS), if sufficiently close to Earth, or alternatively the DSN or ESTRACK networks. Another method could involve an orbit propagation algorithm (Sheikh et al. 2006) together with historical position data. It is assumed that there is a sufficiently accurate time reference onboard the craft providing traceability to terrestrial time scales. Range measurements are obtained in the direction of one or more pulsars between the craft and an inertial reference frame, usually taken to have its origin at the Solar System Barycentre (SSB). To achieve this it is firstly necessary for a pulsar timing model to be available on the craft which gives the TOAs of pulses predicted at the SSB. Secondly, a measurement is required of the TOA of a pulse from a given pulsar. This TOA is then converted to the time the pulse would arrive at the SSB given an initial estimated position of the craft. The unit vector to the pulsar, $\hat{\underline{n}}$, with respect to the SSB is used to achieve this. If the initial estimated position is correct and there are no TOA measurement errors then there will be no time-offset, $\Delta t$, compared to the predicted TOAs at the SSB given by the timing model. However, if the initial estimated position is in error then a non-zero time-offset will be measured corresponding to a position-offset referred to as the 'delta-correction', $\hat{\underline{n}} \cdot \underline{\Delta r} = c\Delta t$, in the direction of the pulsar as shown in Figure 1. Any such discrepancy is used to obtain a corrected spacecraft one-dimensional position estimate in the direction of the pulsar.

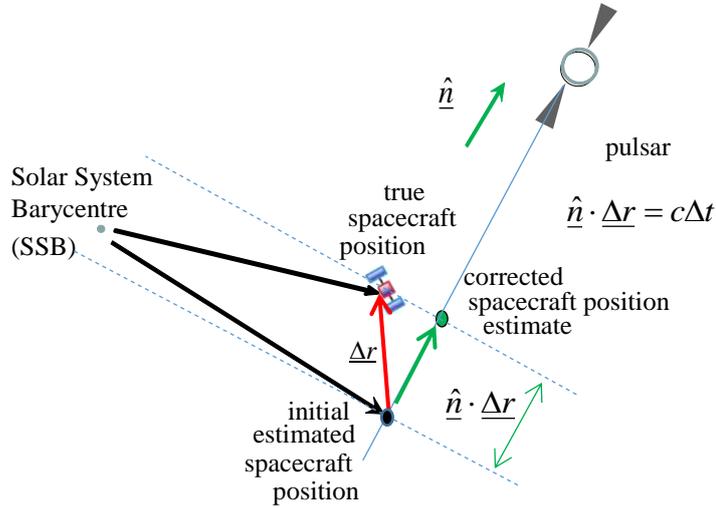

**Fig. 1** Use of delta-correction measurements for estimating the position of a spacecraft in the direction of a pulsar (figure taken from Shemar et al. 2016). The dashed lines represent a given pulse phase of a signal from the pulsar arriving at the true craft position and an initial estimated position at two instants in time separated by an interval $\Delta t$. A measurement of this interval can be used to obtain an estimate of the delta-correction using $c\Delta t$. The green point represents the corrected craft position along the direction of the pulsar.

The advantage of this strategy is that it requires only one pulsar to be observed at a time, which means that the X-ray instrumentation requirements are less complex.

Furthermore, by observing multiple pulsars in sequence it would be possible to derive three-dimensional positioning information, provided the craft motion could be taken account of adequately (Sheikh et al. 2006).

### Absolute navigation

This strategy has some similarities to the processing applied in Global Navigation Satellite Systems (GNSS) such as GPS and offers the potential for periods of autonomous positioning in three dimensions with respect to an inertial reference frame (Sheikh et al. 2006; Sheikh et al. 2007). The advantage of this is that it can navigate and restart without the aid of another method such as DSN. As shown in Figure 2, if there was a significant spacecraft clock time-offset from terrestrial time scales, this could be estimated and corrected by taking measurements of the pulse-phase of a minimum of four pulsars. A minimum of three pulsars would be required if there was a sufficiently accurate time reference available on the craft, which could be obtained using the existing ESTRACK or DSN systems. However, in order to have increased autonomy from the use of existing systems, there would need to be a high-performance atomic clock on-board the craft.

Unlike GNSS signals which include time information, a pulsar signal comprises solely of pulses, with no information regarding the pulse number. Consequently, cycle ambiguities arise that can be resolved using the phase measurements from multiple pulsars, together with knowledge of the unit vector of each pulsar (Sheikh et al. 2006). A priori knowledge, such as an approximate location of the craft within the Solar System and an estimate of the craft velocity together with observations of longer period pulsars may initially be required in order to reduce the parameter search space (Sheikh et al. 2007).

The geometry of the positions of pulsars in the sky is also a factor in navigation performance. For position estimation, a measure of the geometry is the Position Dilution of Precision (PDOP) (Sheikh et al. 2006; Kaplan 1996; Parkinson and Spilker 1996). A lower PDOP value corresponds to a better geometry whereby the pulsars are more widely distributed in the sky. In the case of position and time estimation, the relevant measure of the geometry is the Geometric Dilution of Precision (GDOP). In the case of GPS, typical values of PDOP and GDOP, corresponding to the geometry of satellites in the sky and the user receiver, are between 2 and 3 (Kaplan 1996). In the case of XNAV, although the majority of the pulsars lie near the galactic plane of the Milky Way, some pulsars have positions sufficiently away from the galactic plane to allow good angular separations between a set of three or four. Values of PDOP and GDOP for specific sets of pulsars are given in Table 4, showing that similar values to those obtained in GPS can be achieved in XNAV.

Depending on its navigation uncertainties, absolute navigation would in principle be the most versatile for many types of space missions with potential for greater autonomy. A disadvantage is that it requires simultaneous observations of multiple pulsars which would require multiple detectors, making this strategy more difficult to implement. Whether or not it would be possible to use data taken sequentially for different pulsars using a single detector would depend on the motion of the craft and being able to adequately take account of this (Deng et al. 2013).

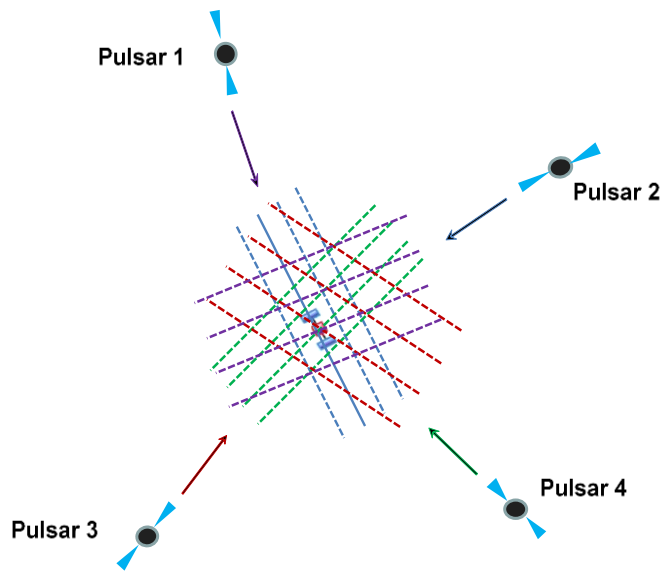

**Fig. 2** Absolute navigation using simultaneous observations of a minimum of four pulsars enabling measurement of the spacecraft three-dimensional position $(x, y, z)$ and the on-board clock time-offset, $t_c$, from terrestrial time scales (figure taken from Shemar et al. 2016). The dashed lines represent candidate lines of position for each pulsar separated by $cP$ in each case and obtained using the measured pulse phases of the four pulsars at a given time.

## 3     X-ray pulsars and XNAV range errors

A description is given here of some X-ray pulsars that may have utility in XNAV. We also provide estimates of spacecraft range error obtained using, firstly, an analytic formula and, secondly, a simulated observation and discuss how they compare.

Although the majority of pulsars were discovered as radio pulsars, ~100 of these are Rotation-Powered X-Ray Pulsars (RP-XRPs) (Becker 2009). As of mid-2012, pulsed X-ray emission has been detected and a pulse profile measured for ~35 of these. They exhibit pulse periods ranging from ~1.5 to 100s of milliseconds (Becker 2009). The Crab pulsar is one of the brightest X-ray sources in the sky, while other known rotation-powered pulsars are typically a thousand times fainter. A relatively small fraction of radio pulsars, and some of the RP-XRPs, can exhibit very good long-term timing stability. These are the 'MilliSecond Pulsars' (MSPs) which have the shortest periods and are relatively weak in strength (Lyne and Smith 2012).

An X-ray pulsar catalogue was compiled, the starting point of which were the tabulations in Becker 2009 which list 89 RP-XRPs. The most detailed information was provided for 35 RP-XRPs for which pulsed X-ray emission profiles were found in the literature. As described fully in Shemar et al. (2016), the catalogue also included data from other databases on pulsar X-ray fluxes, ephemerides, celestial coordinates, X-ray pulse width and fractional pulsed signal. Table 1 summarises parameters for 10 RP-XRPs with detected pulsations that have the lowest range errors. The table includes pulsar period, astrometric position error and estimates of the spacecraft range error contributions for focusing and collimated instrument types. The range error due to the instrument is simply defined as the TOA measurement error for a given pulsar multiplied by the speed of light.

Estimates of the spacecraft range error, based on estimates of the TOA measurement uncertainty, are firstly obtained with a simple analytic formula by using the characteristics of each pulsar, including the X-ray flux, and the proposed instrumentation (e.g. Ray et al. 2008). Table 1 shows range-error estimates obtained this way for focusing- and collimated-type instruments. These values are taken from Shemar et al. (2016) which also describes the formula used and relevant details. They correspond to an observation time, Tobs = $5 \times 10^3$ s, and an instrument effective area, Aeff = 0.005 m$^2$, for the focusing instrument with improved Point Spread Function (PSF) as described in Section 5.

Range error estimates have also been obtained using simulations, as shown in Table 1. They provide more accurate range-error estimates than the analytical method, as they take account of details of the pulse shape of a pulsar and the instrument characteristics. The values taken from Shemar et al. (2016), corresponding to the focusing instrument described in Section 5, are given for the five pulsars PSR B1937+21, B1821-24, J0437-4715, J1012+5307 and B0531+21, and for two integration times of $5 \times 10^3$ and $5 \times 10^4$ s. For each pulsar, the results were obtained using a simulated observation, with data on the pulsed and unpulsed signal, background flux and the instrument response.

For a wide range of pulse-profile shapes and SNR, the values given by the formula and determined by the simulations for the five pulsars agree to within a factor of ~3, as can be seen in Table 1. The estimates from the two approaches show general consistency. The simulations allow a more detailed evaluation of specific cases, for example the effects of low SNR or complex pulse profiles.

| PSR | Type | Pulsar period (s) | Range error – analytical focusing $T_{obs}=5x10^3$ s (km) | Range error – analytical collimated $T_{obs}=5x10^3$ s (km) | Range error – simulated focusing $T_{obs}=5x10^3$ s (km) | Range error – simulated focusing $T_{obs}=5x10^4$ s (km) | Position error (mas) |
|---|---|---|---|---|---|---|---|
| B1937+21 | ms | 0.00155 | 1.2 | 16 | 4.4 | 1.2 | 0.04 |
| B0531+21 | fd | 0.03308 | 1.5 | 1.3 | 2.5 | 0.7 | 3.43 |
| B1821-24 | ms | 0.00305 | 5.9 | 110 | 10.0 | 2.8 | 6.01 |
| J0218+4232 | ms | 0.00232 | 9.1 | 100 | | | 31.12 |
| J0205+6449 | fd | 0.06568 | 45 | 110 | | | 714.49 |
| J0437-4715 | ms | 0.00575 | 83 | 670 | 67 | 16 | 0.05 |
| B0540-69 | fd | 0.05035 | 96 | 160 | | | 69.07 |
| J1012+5307 | ms | 0.00525 | 110 | 2800 | 41 | 23 | 0.48 |
| J0030+0451 | ms | 0.00486 | 120 | 1800 | | | 21.35 |
| B1509-58 | fd | 0.15065 | 180 | 290 | | | 1216.4 |

Table 1: Pulsar parameters and range-error values for 10 rotation-powered X-ray pulsars with the lowest estimated range errors resulting from a focusing instrument, based on the analytic formula. The range-error values are given for the focusing instrument described in Section 5, with $T_{obs} = 5x10^3$ s and $A_{eff} = 0.005$ m$^2$, and for a collimated instrument with the same values of $T_{obs}$ and $A_{eff}$. The key to column headings is: 'Type' is the class of XRP, 'ms' for MSP and 'fd' for field from Becker (2009); the pulsar period (Becker 2009); the 'Range error – analytical' columns show the range error contributions for focusing and collimated instrument types according to the analytic formula and 'Range error – simulated' according to simulations for a focusing instrument for two different observation times; 'Position error' is the pulsar sky (astrometric) position error derived from the ATNF catalogue (Manchester et al. 2005) except for B0531+21 (Lobanov et al. 2011).

# 4 Estimation of navigation and timing uncertainties from simulations

In this section we present estimates of uncertainties for each of the two navigation strategies described in Section 2; delta-correction using a single pulsar and absolute navigation using three or four pulsars. This is done in a similar manner to that described in Shemar et al. (2016). Firstly, we briefly describe the Monte Carlo approaches used to simulate the navigation errors. We then summarise the total uncertainties for PVT estimation using the best-performing pulsars identified for each strategy and for other selected pulsars and pulsar combinations. Lastly, we include some corresponding uncertainty budgets.

The simulations involve a similar approach to that given in Graven et al. (2008) which uses propagation of errors in the small perturbation case for the case of positioning with three pulsars. Taking an ecliptic coordinate system in Cartesian form with its origin at the SSB, the spacecraft $x$, $y$ and $z$ position coordinates and clock-offset $t$ when using four pulsars can be defined as follows

Equation 1

$$\begin{bmatrix} x \\ y \\ z \\ -ct \end{bmatrix} = \begin{bmatrix} x_1 & y_1 & z_1 & 1 \\ x_2 & y_2 & z_2 & 1 \\ x_3 & y_3 & z_3 & 1 \\ x_4 & y_4 & z_4 & 1 \end{bmatrix}^{-1} \begin{bmatrix} \dfrac{cP_1}{2\pi}\phi_1 \\ \dfrac{cP_2}{2\pi}\phi_2 \\ \dfrac{cP_3}{2\pi}\phi_3 \\ \dfrac{cP_3}{2\pi}\phi_4 \end{bmatrix}$$

where $\phi_i$, $\begin{bmatrix} x_i & y_i & z_i \end{bmatrix}$ and $P_i$ are the pulse phase information in radians obtained from the measurements relative to that expected at the SSB, the unit vector and the period of the $i$ th pulsar respectively. The errors $\delta x$, $\delta y$ and $\delta z$ in the spacecraft position coordinates and $\delta t_c$ in the clock-offset are given by

Equation 2

$$
\begin{bmatrix} \tilde{\delta x} \\ \tilde{\delta y} \\ \tilde{\delta z} \\ -c\tilde{\delta t_c} \end{bmatrix} = \begin{bmatrix} x_1 & y_1 & z_1 & 1 \\ x_2 & y_2 & z_2 & 1 \\ x_3 & y_3 & z_3 & 1 \\ x_4 & y_4 & z_4 & 1 \end{bmatrix}^{-1} \begin{bmatrix} \dfrac{cP_1}{2\pi}\Delta\phi_1 \\ \dfrac{cP_2}{2\pi}\Delta\phi_2 \\ \dfrac{cP_3}{2\pi}\Delta\phi_3 \\ \dfrac{cP_4}{2\pi}\Delta\phi_4 \end{bmatrix} - \begin{bmatrix} x_1 & y_1 & z_1 & 1 \\ x_2 & y_2 & z_2 & 1 \\ x_3 & y_3 & z_3 & 1 \\ x_4 & y_4 & z_4 & 1 \end{bmatrix}^{-1} \begin{bmatrix} \Delta x_1 & \Delta y_1 & \Delta z_1 \\ \Delta x_2 & \Delta y_2 & \Delta z_2 \\ \Delta x_3 & \Delta y_3 & \Delta z_3 \\ \Delta x_4 & \Delta y_4 & \Delta z_4 \end{bmatrix} \begin{bmatrix} x \\ y \\ z \end{bmatrix}
$$

where $\begin{bmatrix} \Delta x_i & \Delta y_i & \Delta z_i \end{bmatrix}$ is the error in the unit vector of the $i$ th pulsar derived using the errors in right ascension and declination. The error in the measured phase $\Delta\phi_i$, is mainly a result of the phase measurement error due to the X-ray instrument given by $2\pi\sigma_{TOAi}/P_i$, where $\sigma_{TOAi}$ is the TOA measurement error of the $i$ th pulsar. As described later, in the simulations we also include in $\Delta\phi_i$ a contribution for the pulsar timing model error allowing also for an estimated level of timing noise. $cP_i\Delta\phi_i/2\pi$ is equivalent to the overall range error taken for the $i$ th pulsar. We assume any system level timing errors to be small (see also Section 5 - On-board clocks and timing) compared to the error sources described above. The same approach can be used to obtain velocity errors as described in Shemar et al. (2016).

As can be seen from Table 1, a focusing instrument enables significantly lower range errors than a collimator. Consequently, we present the results of simulations involving four pulsars considering a focusing instrument. The instrument effective area is based on existing technology, i.e. the BepiColombo MIXS-T instrument (Fraser et al. 2010) which has an effective area of 0.005 m$^2$, but with an improved PSF, as described in Section 5. For the five pulsars PSR B1937+21, B1821-24, J1012+5307, J0437-4715 and B0531+21, we use estimates of range errors obtained from the more accurate method of simulated observations, as described in Section 3. An autonomous operation period of 3 months during which time there is no contact with Earth-based systems has been assumed. This sets the interval between two timing model updates during a period of autonomous operation and allows us to take account of the range error contributed by pulsar timing model error. In the case of the Crab pulsar, a much shorter period of order 3 days has been assumed.

Equation 2 enables the simulation of errors for position and clock time-offset, using errors in pulse phase measurements and pulsar positions for four pulsars taken from the pulsar catalogue described in Section 3. Random errors, according to a Gaussian distribution with 1 sigma given by the uncertainty in the relevant parameter, have been generated for each pulsar as inputs to Equation 2 in order to determine $\delta x$, $\delta y$, $\delta z$ and $\delta t_c$. This is repeated 100 times in order that the distribution of output errors in each case can be used to estimate a representative value for the 1-sigma uncertainty. A similar method is used to derive 1-sigma uncertainties in all other PVT estimation cases, including for absolute navigation using three pulsars and delta-correction using a single pulsar.

The navigation and timing uncertainty budgets comprise mainly two components. The first is largely due to the TOA measurement uncertainties arising from the instrument, given by the first term on the right-hand side of Equation 2, with a contribution also from the pulsar timing model uncertainty. The second is due to the pulsar position uncertainties on the sky, given by the second term on the right-hand side of Equation 2. This component leads to the craft position and clock time-offset uncertainties increasing linearly with range from the SSB, whilst the velocity and time-drift rate uncertainties increase linearly with velocity.

Shemar et al. (2016) presented uncertainty budgets for the pulsars that give the lowest positioning uncertainties for each navigation strategy in the case of a craft with position coordinates given by $x$=30 AU, $y$=0 AU and $z$=0 AU and with velocity vector components $v_x$=30 kms$^{-1}$, $v_y$=0 kms$^{-1}$ and $v_z$= 0 kms$^{-1}$. The craft position in this case is equivalent to a range of 30 AU from the SSB in the direction of zero degrees ecliptic longitude and latitude. PSR B1937+21 was used for the single pulsar case with the delta-correction strategy, PSR B1937+21, B1821-24 and J0437-4715 for absolute navigation using three pulsars and PSR B1937+21, B1821-24, J1012+5307 and J0437-4715 for absolute navigation using four pulsars. As the Crab pulsar may be of particular interest, due to its relative brightness, results were also shown for the single pulsar case of the Crab pulsar and for the three pulsar-set PSR B1937+21, B0531+21, J0437-4715, which gives the lowest uncertainty of any three pulsar-set that includes the Crab pulsar. In Table 2, we present results using the above single pulsar and three pulsar-set cases, but at a range of 1 AU from the SSB within the ecliptic plane instead of 30 AU as in the case of Shemar et al. (2016). These firstly apply to a craft with position coordinates given by $x$=1 AU, $y$=0 AU and $z$=0 AU and velocity vector components $v_x$=30 kms$^{-1}$, $v_y$=0 kms$^{-1}$ and $v_z$= 0 kms$^{-1}$. Secondly, to show the potential variation with craft position in the ecliptic plane and craft velocity vector, we also present those (as given in parentheses) for a craft with position coordinates $x$=0 AU, $y$=1 AU and $z$=0 AU and velocity vector components $v_x$=0 kms$^{-1}$, $v_y$=30 kms$^{-1}$ and $v_z$= 0 kms$^{-1}$. In Table 3, we similarly present the uncertainty budgets for the above-mentioned four-pulsar-set for absolute navigation.

The PVT uncertainties are summarized in Tables 4 and 5 for craft ranges of 30 and 1 AU from the SSB, while Table 4 also gives the PDOP and GDOP values for the three and four pulsar cases respectively. The latter values show that similar values of PDOP and GDOP can be achieved in XNAV as in GPS (see also Section 2 – Absolute navigation). It should be noted that the best-performing pulsar-sets given in Tables 2 to 5 correspond to those giving the lowest craft positioning uncertainties at ecliptic coordinates given by $x$=30 AU, $y$=0 AU and $z$=0 AU. Where uncertainties are presented for a range of 1 AU from the SSB, other pulsar-sets may have similar or marginally lower uncertainties.

The simulations are applicable to scenarios related to interplanetary navigation, which accounts for the majority of deep space missions. Furthermore, due to their dependence on the exact values of the input data used for each pulsar, the results should be taken as order of magnitude estimates.

| Uncertainty Source | Position Uncertainty Contribution (km) | | Velocity Uncertainty Contribution (ms$^{-1}$) | | Position Uncertainty Contribution (km) | | Velocity Uncertainty Contribution (ms$^{-1}$) | |
|---|---|---|---|---|---|---|---|---|
| | $T_{obs} =$ 5x10$^4$ s | $T_{obs} =$ 5x10$^3$ s | $T_{obs} =$ 5x10$^4$ s | $T_{obs} =$ 5x10$^3$ s | $T_{obs} =$ 5x10$^4$ s | $T_{obs} =$ 5x10$^3$ s | $T_{obs} =$ 5x10$^4$ s | $T_{obs} =$ 5x10$^3$ s |
| | PSR B1937+21 | | | | PSR B0531+21 | | | |
| Instrument and timing model | 1.4 | 4.6 | 0.03 | 1 | 3.7 | 5.5 | 0.07 | 1 |
| Pulsar position | 0.02(0.02) | 0.02(0.02) | 5(5)x10$^{-6}$ | 5(5)x10$^{-6}$ | 2(0.2) | 2(0.2) | 500(50)x10$^{-6}$ | 500(50)x10$^{-6}$ |
| Total uncertainty | 1.4(1.4) | 4.6(4.6) | 0.03(0.03) | 1(1) | 4.5(3.7) | 6(5.5) | 0.07(0.07) | 1(1) |
| | PSR B1937+21, PSR B1821-24, PSR J0437-4715 | | | | PSR B1937+21, PSR B0531+21, PSR J0437-4715 | | | |
| Instrument and timing model | 30 | 120 | 0.6 | 25 | 30 | 110 | 0.6 | 20 |
| Pulsar position | 0.6(0.3) | 0.6(0.3) | 0.1(0.05)x10$^{-3}$ | 0.1(0.05)x10$^{-3}$ | 3(0.3) | 3(0.3) | 0.7(0.07)x10$^{-3}$ | 0.7(0.07)x10$^{-3}$ |
| Total uncertainty | 30(30) | 120(120) | 0.6(0.6) | 25(25) | 30(30) | 110(110) | 0.6(0.6) | 20(20) |

Table 2: Spacecraft position and velocity uncertainty budgets, firstly for the delta-correction method using a single pulsar for the cases of PSR B1937+21 and B0531+21, corresponding to a craft located at ecliptic coordinates $x$=1 AU, $y$=0 AU and $z$=0 AU and with velocity vector components $v_x$=30 kms$^{-1}$, $v_y$=0 kms$^{-1}$ and $v_z$= 0 kms$^{-1}$. Secondly, these are given for two three-pulsar-sets for the case of absolute navigation. The uncertainties for a craft located at $x$=0 AU, $y$=1 AU and $z$=0 AU and with velocity vector components $v_x$=0 kms$^{-1}$, $v_y$=30 kms$^{-1}$ and $v_z$= 0 kms$^{-1}$ are also shown (in parentheses).

| Uncertainty Source | Position Uncertainty Contribution (km) | | Time-Offset Uncertainty Contribution (s) | | Velocity Uncertainty Contribution (ms$^{-1}$) | | Time Drift Rate Uncertainty Contribution (ss$^{-1}$) | |
|---|---|---|---|---|---|---|---|---|
| | $T_{obs}=$ $5\times10^4$ s | $T_{obs}=$ $5\times10^3$ s | $T_{obs}=$ $5\times10^4$ s | $T_{obs}=$ $5\times10^3$ s | $T_{obs}=$ $5\times10^4$ s | $T_{obs}=$ $5\times10^3$ s | $T_{obs}=$ $5\times10^4$ s | $T_{obs}=$ $5\times10^3$ s |
| | PSR B1937+21, PSR B1821-24, PSR J1012+5307 and PSR J0437-4715 | | | | | | | |
| Instrument and timing model | 30 | 80 | $40\times10^{-6}$ | $100\times10^{-6}$ | 0.6 | 15 | $1\times10^{-9}$ | $20\times10^{-9}$ |
| Pulsar position | 0.7(0.5) | 0.7(0.5) | $0.3(0.2)\times10^{-6}$ | $0.3(0.2)\times10^{-6}$ | $0.2(0.1)\times10^{-3}$ | $0.2(0.1)\times10^{-3}$ | $1(0.5)\times10^{-13}$ | $1(0.5)\times10^{-13}$ |
| Total uncertainty | 30(30) | 80(80) | $40(40)\times10^{-6}$ | $100(100)\times10^{-6}$ | 0.6(0.6) | 15(15) | $1(1)\times10^{-9}$ | $20(20)\times10^{-9}$ |

Table 3: Spacecraft position, clock time-offset, velocity and time drift rate uncertainty budgets corresponding to a craft located at ecliptic coordinates $x$=1 AU, $y$=0 AU and $z$=0 AU and with velocity vector components $v_x$=30 kms$^{-1}$, $v_y$=0 kms$^{-1}$ and $v_z$= 0 kms$^{-1}$, using a particular four-pulsar-set (see main text for choice of pulsars) in the case of absolute navigation. The uncertainties for a craft located at $x$=0 AU, $y$=1 AU and $z$=0 AU and with velocity vector components $v_x$=0 kms$^{-1}$, $v_y$=30 kms$^{-1}$ and $v_z$= 0 kms$^{-1}$ are also shown (in parentheses).

| Navigation strategy & pulsars | PDOP/ GDOP | Parameter | 30 AU | | 1 AU | |
|---|---|---|---|---|---|---|
| | | | $T_{obs}= 5 \times 10^4$ s | $T_{obs}= 5 \times 10^3$ s | $T_{obs}= 5 \times 10^4$ s | $T_{obs}= 5 \times 10^3$ s |
| (i) Absolute navigation using three pulsars | PDOP=2.9 | Position(km) | 35(30) | 120(120) | 30(30) | 120(120) |
| | | Velocity(ms$^{-1}$) | 0.6(0.6) | 25(25) | 0.6(0.6) | 25(25) |
| PSR B1937+21, B1821-24, J0437-4715 | | | | | | |
| (iii) Absolute navigation using three pulsars | PDOP=2.6 | Position(km) | 100(30) | 140(110) | 30(30) | 110(110) |
| | | Velocity(ms$^{-1}$) | 0.6(0.6) | 20(20) | 0.6(0.6) | 20(20) |
| PSR B1937+21, B0531+21, J0437-4715 | | | | | | |
| (iv) Absolute navigation using four pulsars | GDOP=2.6 | Position(km) | 35(35) | 80(80) | 30(30) | 80(80) |
| | | Clock time-off-set(s) | $40(40) \times 10^{-6}$ | $100(100) \times 10^{-6}$ | $40(40) \times 10^{-6}$ | $100(100) \times 10^{-6}$ |
| PSR B1937+21, B1821-24, J1012+5307, J0437-4715 | | Velocity(ms$^{-1}$) | 0.6(0.6) | 15(15) | 0.6(0.6) | 15(15) |
| | | Clock time-drift rate(ss$^{-1}$) | $1(1) \times 10^{-9}$ | $20(20) \times 10^{-9}$ | $1(1) \times 10^{-9}$ | $20(20) \times 10^{-9}$ |

Table 4: A summary of the PVT uncertainties for specific cases of absolute navigation using three and four pulsars. These are given for a spacecraft located in the ecliptic plane at a distance of 30 AU and 1 AU from the SSB in the direction of zero degrees ecliptic longitude and latitude, corresponding to ecliptic coordinates $x$=30 AU, $y$=0 AU, $z$=0 AU and $x$=1 AU, $y$=0 AU, $z$=0 AU respectively. The velocity vector components in each case are $v_x$=30 kms$^{-1}$, $v_y$=0 kms$^{-1}$ and $v_z$= 0 kms$^{-1}$. The uncertainties for a craft located at coordinates $x$=0 AU, $y$=30 AU, $z$=0 AU and $x$=0 AU, $y$=1 AU and $z$=0 AU and with velocity vector components $v_x$=0 kms$^{-1}$, $v_y$=30 kms$^{-1}$ and $v_z$= 0 kms$^{-1}$ are also shown (in parentheses). Values are also given for the PDOP in the cases of using three pulsars, and GDOP for the four pulsar case.

| Navigation strategy & pulsars | Parameter | 30 AU | | 1 AU | |
|---|---|---|---|---|---|
| | | $T_{obs}$= 5x10$^4$ s | $T_{obs}$= 5x10$^3$ s | $T_{obs}$= 5x10$^4$ s | $T_{obs}$= 5x10$^3$ s |
| (v) Delta-correction using a single pulsar with PSR B1937+21 | Position(km)(in direction of pulsar) | 1.6(1.5) | 4.7(4.7) | 1.4(1.4) | 4.6(4.6) |
| | Velocity(ms$^{-1}$) (in direction of pulsar) | 0.03(0.03) | 1(1) | 0.03(0.03) | 1(1) |
| (vi) Delta-correction using a single pulsar with PSR B0531+21 | Position (in direction of pulsar) | 70(10) | 70(10) | 4.5(3.7) | 6(5.5) |
| | Velocity(ms$^{-1}$) (in direction of pulsar) | 0.07(0.07) | 1(1) | 0.07(0.07) | 1(1) |

Table 5: A summary of the navigation uncertainties for the single pulsar cases of PSR B1937+21 and PSR B0531+21. These are given for a spacecraft located in the ecliptic plane at a distance of 30 AU and 1 AU from the SSB in the direction of zero degrees ecliptic longitude and latitude, corresponding to ecliptic coordinates $x$=30 AU, $y$=0 AU, $z$=0 AU and $x$=1 AU, $y$=0 AU, $z$=0 AU respectively. The velocity vector components in each case are $v_x$=30 kms$^{-1}$, $v_y$=0 kms$^{-1}$ and $v_z$= 0 kms$^{-1}$. The uncertainties for a craft located at coordinates $x$=0 AU, $y$=30 AU, $z$=0 AU and $x$=0 AU, $y$=1 AU and $z$=0 AU and with velocity vector components $v_x$=0 kms$^{-1}$, $v_y$=30 kms$^{-1}$ and $v_z$= 0 kms$^{-1}$ are also shown (in parentheses).

# 5    X-Ray Technology

Determination of a spacecraft's position in space by observing pulsars requires instrumentation able to unambiguously determine the phase of the pulsed signal against an absolute timing reference. This places a number of constraints on the instrumentation used to measure the signals from the pulsar, which must be designed to maximize the desired signal while minimizing background sources from the sky and the local charged particle environment close to the craft. Practical considerations, imposed by the necessities of a deep-space mission, further constrain the hardware that can be used, placing strict limits on the size and mass of the instrumentation. Here we discuss the merits of various instrumentation designs and how they can achieve the navigation performances described above.

There are two key instrument families that are applicable to XNAV, collimated systems an imagers. The requirements of the two systems are different, but a strong emphasis is given to imaging instrumentation due to its greatly reduced background for a given collecting area (leading to better navigation performance). This section will show that maturing lightweight optics technology can provide a practical method of implementing a navigation subsystem within a realistic resource envelope on a deep-space craft.

## System requirements

The requirements of an XNAV instrument are very similar to existing science concepts, in that the device must detect pulsar signals, while minimising error sources such as sky background, particle induced events in the detectors, internal background and spurious timing modulations e.g. due to the pointing stability.

An example of the technical requirements for an XNAV instrument can be summarised as:

- High time resolution (<1 µs, goal <300 ns)
- High collecting area (~50 cm$^2$ @1 keV for imager)
- Energy range ~0.5-8 keV
- Low background noise
- Sufficiently accurate on-board time

It is straightforward to derive a hugely capable instrument based on these requirements without consideration of the context of the missions for which it would be used.

A deep space payload is likely to remain highly mass-constrained in the next few decades and, therefore, for the system to be realistic it is important to place similar resource constraints on the navigation system to the scientific instrument payload. Here, the BepiColombo Mercury Imaging X-ray Spectrometer (Fraser et. al. 2010), which will be the first imaging X-ray instrument to fly on a deep-space mission, was used as a baseline and perturbations to its design used to optimise a concept design for an XNAV system.

## Optics technologies

The optic design is fundamental to the ability to realise a low-mass X-ray telescope for practical use on a deep-space craft. A number of optic technologies are capable of providing instrumentation for XNAV, but because of the self-imposed resource constraints considered here, the most promising is based on square pore MicroChannel-Plates (MCPs). These can be arranged in a number of geometries to provide the necessary focusing. These are reviewed in detail by R. Willingale in this Volume. Below we state the specific advantages and limitations of these designs for XNAV applications.

*Wolter I optics* – It is possible to approximate to the complex paraboloidal and hyperboloidal surfaces of the Wolter I design with short, straight channels by slumping the MCP such that the pores all point to the centre of a sphere. Sequential reflections off two MCPs of different slump radii bring the rays to a point-like focus. The design is well proven by the BepiColombo MIXS instrument, but is difficult and expensive to realise.

*Narrow-field lobster optic* – Lobster imaging, as reported by Angel (1979), relies on the fact that every line-of-sight has equivalent optic structures projected towards it, i.e. there is no preferred axis for the optic. Moving away from this model by profiling the thickness of the optic structures, it is possible to generate a large effective area in a given axis at the cost of altering the off-axis vignetting function (see R. Willingale paper in this Volume). This type of optic is very well matched to the XNAV requirements, and is much simpler to implement than the other geometries discussed.

*Kirkpatrick-Baez (KB) optics* – This is somewhat simpler than the Wolter system. However, a KB optic also possesses a number of the drawbacks in cost and complexity, when compared to the simple lobster geometries described above. An MCP approximation to the KB geometry relies on two square pore, square packed optics being presented to the beam in sequence. Both are slumped to a cylindrical figure, such that the pores point to a line centred on the axis of the cylinder. Unlike the Wolter system, both MCPs have the same curvature, simplifying manufacture and removing the requirement to force a rectilinear array onto a spherical surface (as is the case for Lobster imaging). Therefore, in principle, this design can lead to an optic with fewer aberrations and therefore a better PSF.

Figure 3 shows the first experimental data from a cylindrically curved MCP. This image demonstrates that fabrication of the MCP structures required for a KB system is verified. Data were taken with a broad continuum of X-rays up to 20 keV with the W-L line at 8.9 keV superimposed (as generated using a Phillips PW1730/10 100 kV X-ray generator with a PW2184/00 tungsten anode X-ray tube). The imaging quality is found to be ~ 3.6 arcmin Full Width at Half Maximum (FWHM) for the optic which is cylindrically curved with a radius of 3 m. The focused rays are evident as a vertical line in the image. A fundamental performance estimation and detailed optic design remains to be done, as does experimentally verifying the correct operation of a tandem stack of two perpendicular MCPs to perform true imaging. However, this optic type is very promising as it could offer significantly better focusing with an angular resolution ~3 times smaller than realised for MIXS-T (and hence ~9x lower background as the background scales with the PSF area). As the technological maturity is low, more work is required for a KB system to become a viable option for future instrument design studies, however, the XNAV performance estimates presented elsewhere in this paper are based on an instrument with this improved PSF.

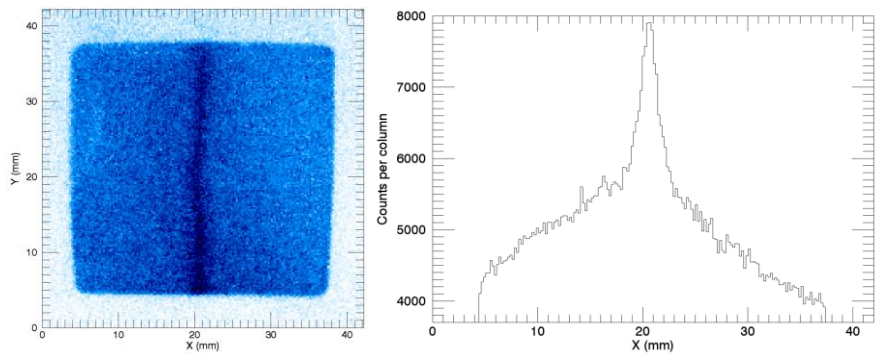

**Fig. 3** The first experimental data obtained with a cylindrically slumped MCP, showing the expected line focus is realised. Left, detector image. Right, profile of image integrated along the columns, showing the narrow line from the focused X-rays. The FWHM of the focus is ~3.6 arcminutes implying a factor 3 improvement in uni-directional focusing compared to the MIXS-T optic.

## Detector technologies

There is a limited subset of detectors capable of meeting the requirements of an XNAV system, some of these will be discussed in detail elsewhere in these proceedings (see e.g. Meidinger et al. paper in this Volume).

Gas counters offer good timing resolution, good efficiency (above energies affected by the window) and high technology maturity. However, they are large devices, which (in general) need anticoincidence shielding to reduce background and a gas supply (if the window is thin enough to allow gas constituency changes as gas escapes through the window). As such, they are not the most optimised solution for XNAV, which would require the subsystem to be resilient for the many years necessary to complete the mission.

MCP detectors are by far the highest time resolution device considered here, allowing fundamental timing limited to ~10 ps. However, MCPs have a significantly lower quantum detection efficiency (~30 %) relative to other technologies, meaning that the optic would have to provide ~3 times the area to achieve the same signal level. Other drawbacks include, very limited energy resolution (no background rejection on energy grounds), internal background from radioisotopes in the glass and added complexity. The accuracy in timing offered by MCPs is not necessary to ensure that pulse time of arrival measurements from the pulsars are limited by the astrophysics of the pulsar (rather than the detector/electronics), hence MCPs are not considered a good match to XNAV requirements.

Silicon Drift Detectors (SDDs) offer ~10 µs time resolution in large arrays (e.g. Barret et al. 2010) and ~0.5 µs in single pixel format, along with good spectral resolution and high quantum efficiency. They offer very good energy resolution, low energy thresholds of <200 eV and high quantum efficiency. These devices are an excellent match for XNAV and will be used on the Neutron star Interior Composition Explorer (NICER) experiment (see Gendreau paper in this Volume).

Avalanche PhotoDiodes (APDs) are compact, lightweight devices which offer ~100 % quantum efficiency and ~few ns timing resolution. Device sizes are limited to a maximum of ~5-10 mm diameter, which is sufficient to make them a good match for an XNAV system. A large APD device is reported as an X-ray detector by Ikagawa (2005), demonstrating good technology maturity. The major drawback of an APD system is that it has degraded spectroscopy and a higher low energy threshold than SDDs (e.g. Kataoka et al. 2004). This is because APDs have internal gain and therefore increased noise due to the statistics of the charge multiplication process. Literature suggests that it is possible to generate a low energy threshold ~0.5 keV by cooling the APD (e.g. Lynch et. al. 1996, or Kataoka et.al. 2005)

A full trade-off between SDDs and APDs is outside the scope of this study but both are good options for XNAV systems.

## Other considerations

### Source acquisition and pointing requirements

To acquire the source in the XNAV instrument, the satellite pointing stability and pointing knowledge must be such that the source arrives within the instrument field of view and the required fraction of the PSF lies on the detector and remains there for the duration of the observation. A typical pointing accuracy and stability requirement may be of order ±30 arcsec to avoid the loss of signal from the edge of the detector. This is not considered a major challenge for a typical attitude and orbit control system. However, deep space missions in the future will likely rely on low thrust propulsion systems, meaning that the pointing of the XNAV instrument to its targets is likely to require independent pointing control as the thrust axis of the satellite must remain fixed. When considering an independently pointed instrument on a satellite platform, an error stack arises including; the satellite attitude knowledge and stability, the thermal and mechanical tolerances and stability of the pointing mechanism and the relationship of the instrument field of view relative to the star tracker coordinate system.

*On-board clocks and timing*

Maintaining onboard timing accuracy is a key driver for minimizing the required number of observed pulsars in the case of absolute navigation from four to three. Time transfer to terrestrial time scales could be achieved by periodic calibration using the DSN or ESTRACK networks. However, greater autonomy from such systems could be achieved by using a sufficiently accurate atomic clock on-board the craft. Such a device would, for example, need to maintain an error with respect to a terrestrial time standard (e.g. TAI) of lower than 300 ns over the duration of the deep space mission. In the next few years, it is possible that a clock with sufficient stability could be demonstrated in space, for example, the NASA Deep Space Atomic Clock (DSAC) (Tjoelker et al. 2011).

## Instrument Configuration

Based on the consideration of the available technologies, a baseline payload can be envisaged which would consist of a low mass MCP optic similar to those used on the MIXS telescope. The simplest solution would be the Lobster design, the highest maturity would be the Wolter design and the best suited, though least mature technology is likely to be the Kirkpatrick-Baez design. An SDD detector similar to those used on NICER and a high stability atomic clock, for example, the NASA DSAC would complete the concept payload.

Shemar et al. (2016) provides a detailed breakdown of possible technology options, and shows a parametric model of the payload capability as a function of mass and focal length. The conclusion is that a highly capable navigation performance can be achieved in a small, compact instrument with mass ~12 kg, volume ~1000x250x250 mm$^3$ and power ~ 15-20 W. Smaller, lower mass options exist for specific navigation scenarios. Further work is needed to assess the spacecraft systems needed to support this kind of device, e.g. the pointing and thermal control systems

# 6 Discussion

**Comparison of performance between XNAV and DSN**

At a range of 30 AU and for Tobs = 5x10$^4$ s, three-dimensional XNAV positioning uncertainties for an improved PSF version of the BepiColombo-MIXS instrument given in Table 4 are generally ~one order of magnitude greater than those described in Section 1 for the DSN, although with the potential for being somewhat lower in the direction of PSR B1937+21 as given in Table 5. The XNAV positioning uncertainties are generally of the same order as that of the ESTRACK system at a range of 30 AU, although more than an order of magnitude lower in the direction of PSR B1937+21. A key advantage for XNAV is that, unlike the use of the ESTRACK or DSN systems for navigation, it could allow a greater level of spacecraft autonomy because it requires significantly less communication with Earth-based systems.

**XNAV performance within the ecliptic plane**

It can be seen that for a given value of $T_{obs}$ = 5x10$^4$ s or 5x10$^3$ s, the XNAV performance is generally of the same order at 30 AU and 1 AU from the SSB within the ecliptic plane. This is because the range error contribution due to the instrument is a significant part of the uncertainty budget. The three-pulsar-set given by PSR B1937+21, B1821-24, J0437-4715 shows the lowest position uncertainties and one of the lowest velocity uncertainties for $T_{obs}$ = 5x10$^4$ s out of all the pulsar combinations considered. For a typical distance of Mars from the SSB, 1.5 AU, the total uncertainty is ~30 km after observing each pulsar for 5x10$^4$ s. Furthermore, it is found that the uncertainties do not vary substantially with ecliptic longitude, apart from for the cases that include the Crab pulsar, which has a relatively high astrometric position error.

By assuming that the motion of the craft can be adequately accounted for, the XNAV uncertainties due to the instrument would reduce by $\sqrt{N}$ where N is the number of observations. However, in practice this will be limited by the errors in the trajectory models including due to the effects of tertiary bodies in the Solar System and solar radiation pressure over long time intervals (Deng et al. 2013).

**XNAV in combination with ESTRACK / DSN**

In the case of using a single pulsar, uncertainties of ~5 and ~1.5 km may be achieved in the direction of PSR B1937+21 with $T_{obs}$ = 5x10$^3$ s (~1 hour) and 5x10$^4$ s (~10 hours) respectively for a distance of up to 30 AU from the SSB. The DSN and ESTRACK systems would require up to 8 hours for providing a craft with position information in this scenario. XNAV has the potential to provide more accurate position information along the direction of PSR B1937+21. Depending on the geometry of the particular scenario, this could then be combined with the information from the ESTRACK or DSN system to enable reduced position uncertainties in the plane of the sky (Graven et al. 2008).

**Performance for autonomous operation**

As mentioned already, at a typical distance of Mars from the SSB, 1.5 AU, the three-pulsar-set PSR B1937+21, B1821-24, J0437-4715 would enable an uncertainty of ~30 km after observing each pulsar for $5x10^4$ s. These uncertainties could be generally achieved without the need to communicate with Earth to update the pulsar's timing model for up to ~3 months. A disadvantage of using the set PSR B1937+21, B0531+21 and J0437-4715 would be the need for much higher timing model updates, for example daily, to be transmitted to the craft using the ESTRACK or the DSN systems due to the much higher timing noise and rate of glitches (Lyne and Smith 2012) exhibited by the Crab pulsar. Furthermore, caution is also required as there is a small probability of a glitch occurring in one of the pulsars and this may result in the accuracy of the timing model very rapidly degrading. No glitches have so far been observed in any MSP apart from PSR B1821-24 for which a microglitch has once been observed (Cognard and Backer 2004).

**Instrumentation challenges**

The implementation of an X-ray instrument allowing adequate simultaneous measurements of multiple pulsars on a craft would be a significant challenge. It may be more realistic to consider an instrument that observes pulsars sequentially. For this scenario, the observation times, $T_{obs}$, given in Section 4 for absolute navigation using three and four pulsars would need to be multiplied by three and four respectively. Pulsar position geometry is also a factor in navigation performance, as described in Section 2. Ideally the pulsars should be as widely distributed in the sky as possible.

It is found that similar position and velocity uncertainties can be achieved using three pulsars together with an accurate atomic clock instead of using four pulsars. The need to observe only three pulsars as opposed to four may mean it is possible to have a simpler instrument design in the case where multiple pulsars are to be observed simultaneously.

The instrumentation needs of an XNAV system have been shown in Section 5 to be compatible with current technologies. Significant improvements in performance and increased spacecraft autonomy are expected from the adoption of new technology in the next decade. Potential technological improvements exist in all three of the major subsystems considered; the optic, detector and the craft on-board timing. Improving the optic reduces background noise and offers lower navigation uncertainties. Improving the detector system e.g. to generate very high time resolution would allow improved estimates of a pulse TOA. Although a time resolution of <1 μs may be achieved with current technology, 100 ns may be achievable in the next decade. Finally, improvements in the stability of atomic clocks for space applications will in future offer the possibility of reducing the number of pulsars that need to be observed to allow autonomous three-dimensional navigation from four to three.

## Use of radio observations

It has been proposed that radio observations of pulsars may require a large radio antenna (Becker et al. 2013; A. Jessner in this Volume). A practical solution would need to be identified in order to exploit radio observations in combination with or as an alternative to X-ray observations, considering the resource requirements of a conventional deep space mission.

# 7    Conclusions

A summary of the results in this paper is given below

- A key advantage of XNAV is that, unlike the use of the ESTRACK or DSN systems for navigation, it could allow a greater level of spacecraft autonomy because it requires significantly less communication with Earth-based systems.

- The Mercury Imaging X-ray Spectrometer (MIXS) instrument, due to be launched on the ESA/JAXA BepiColombo mission to Mercury in 2018, is an example of an instrument that may be further developed as a practical telescope to enable XNAV on a conventional deep space mission.

- Using an instrument such as above, PSR B1937+21 may be used to achieve craft positioning uncertainties of ~5 and ~1.5 km in the direction of the pulsar with an observing time $T_{obs} = 5x10^3$ s and $5x10^4$ s respectively for a distance of up to 30 AU from the SSB.

- For a distance of up to 30 AU from the SSB, the three-pulsar-set PSR B1937+21, B1821-24, J0437-4715 would enable a craft positioning uncertainty of ~35 km after observing each pulsar for $5x10^4$ s. A lower uncertainty may be achieved, for example, by use of extended observations or, if feasible, by use of a larger instrument. An atomic clock of sufficient timing stability would allow a greater level of spacecraft autonomy.

- When using three or four pulsars, the geometry of their positions in the sky is also a factor in navigation performance. Values of PDOP and GDOP have been given for a small number of specific sets of three and four pulsars, including those that give the lowest positioning uncertainties. These show that similar values to those obtained in GPS can be achieved in XNAV.

- Possible options for future developments in terms of simpler, lower-cost Kirkpatrick-Baez optics have been discussed, in addition to the principal design and development challenges that must be addressed in order to realise an operational XNAV system.

- Radio observations of pulsars could in principle be used in combination with or as an alternative to X-ray. A practical solution would need to be identified in order to achieve these on a conventional deep space mission considering the resource constraints.

## Acknowledgments

The work reported here is based on earlier work performed primarily under European Space Agency Contract No. 4000105938/12/NL/KML. Microchannel plate optics have been developed by Photonis SAS, Brive France, in collaboration with the University of Leicester funded by ESA contracts. We acknowledge the assistance of Andrew Lamb in the early stages. We are grateful to John Davis at the National Physical Laboratory for constructive comments during preparation of the manuscript.